\documentclass[runningheads]{llncs}
\usepackage{graphicx}
\begin{document}
\title{k-Maximum Subarrays for Small k: Divide-and-Conquer made simpler}
%
%
\author{Ovidiu Daescu \and
Hemant Malik}
\authorrunning{Ovidiu Daescu and Hemant Malik}
%
\institute{University of Texas at Dallas, Richardson TX 75080, USA \\
\email{\{daescu, malik\}@utdallas.edu}}
\maketitle              
\begin{abstract}
Given an array \textit{A} of \textit{n} real numbers, the maximum subarray problem is to find a contiguous subarray which has the largest sum. The \textit{k}-maximum subarrays problem is to find \textit{k} such subarrays with the largest sums. For the 1-maximum subarray the well known divide-and-conquer algorithm, presented in most textbooks, although suboptimal, is easy to implement and  can be made optimal with a simple change that speeds up the combine phase. On the other hand,  the only known divide-and-conquer algorithm for $k > 1$, that is efficient for small values of $k$, is difficult to implement, due to the intricacies of the combine phase. In this paper we give a 
divide-and-conquer solution for the \textit{k}-maximum subarray problem 
that simplifies the combine phase considerably while preserving the overall running time.

In the process of designing the combine phase of the algorithm we provide a simple, sublinear, $O(\sqrt{k} \log^{3} k)$ time algorithm, for finding the $k$ largest sums of $X + Y$, where $X$ and $Y$ are sorted arrays of size $n$ and $k \le n^2$. The $k$ largest sums are implicitly represented, and can be enumerated with an additional $O(k)$ time. 

Unlike previous solutions, that are fairly complicated and sometimes difficult to implement, ours rely on simple operations such as merging sorted arrays, binary search, and selecting the $k$-th smallest number in an array. We have implemented our algorithms and report excellent performance as compared to previous results.

\end{abstract}
\section{Introduction}
The well known problem of finding the maximum sum (contiguous) subarray of a given array of real numbers has been used in various applications and received a lot of attention over time. Some of the applications are in data mining~\cite{agrawal1993mining,fukuda1996data}, pattern recognition~\cite{grenander1978pattern}, and image processing and communication~\cite{bentley1984programming}. 

Given an array \textit{A} of \textit{n} real numbers and an integer \textit{k}, such that $1 \leq k \leq n(n+1)/2$, the k-maximum subarrays problem is to find \textit{k} contiguous subarrays with the largest sums (not necessarily in sorted order of the sums). If \textit{k}=1, Kadane's algorithm~\cite{bentley1984algorithm} solves the maximum subarray problem in O($n$) time using an iterative method. On the other hand, the well known divide and conquer algorithm~\cite{cormen2009introduction}, found in virtually all algorithms textbooks, has a suboptimal O($n \log n$) running time. 
An $O(n)$ time divide-and-conquer algorithm is briefly presented in~\cite{bengtsson2007ranking}.

For $k > 1$, Bengtsson and Chen~\cite{bengtsson2006efficient} presented an algorithm that takes time $O(\min\{k + n \log^{2}n, n \sqrt{k}\}$), where the second term, $O(n \sqrt{k})$, comes from a divide-and-conquer solution. 
That divide-and-conquer algorithm is difficult to implement, due to the intricacies of the combine phase.


In this paper we propose a competitive, $O(n \sqrt{k})$ time divide and conquer solution to find the \textit{k}-maximum subarrays, which is optimal for $k=O(1)$ and $k=O(n^2)$. Our algorithm is simpler than the one in~\cite{bengtsson2006efficient} due to a more direct way of performing the combine phase. Specifically, the combine phase we propose is itself a simple recursive procedure. To this end, we also address the following subproblem: \textit{Given two sorted arrays of real numbers, $X$ and $Y$, each of size $n$, let $S$ be the set $S = \{(x,y)$ $ | x \in X$ and $y \in Y\}$, with the value of each pair in the set defined as $Val(x,y)$ = $x + y$. Find the \textit{k} pairs from \textit{S} with largest values.} This problem is closely related to the famous pairwise sum ($X+Y$) problem~\cite{frederickson1982complexity,frederickson1984generalized}, that asks to sort all pairwise sums. 
Our main contribution is a sublinear, $O(\sqrt{k} \log^{3} k)$ time algorithm, 
for finding the $k$ largest sums of $X + Y$. 
The $k$ largest sums are implicitly represented, and can be enumerated with an additional 
$O(k)$ time. 
A key feature of our solution is its simplicity, compared to previous 
algorithms~\cite{frederickson1982complexity,frederickson1984generalized}, that find and report the $k$ largest sums in $O(k + \sqrt{k})$ time.
Our algorithm 
is very simple, using only operations such as merging sorted arrays, binary search, and selecting the $k$-th largest number of an array.

We have implemented our main algorithms in JAVA and performed extensive experiments on macOS High Sierra with 3.1 GHz intel i5 processor and 8 GB of RAM, reporting excellent performance. For example, on random arrays of size $10^6$, with $k=10^6$, we can find the $k$ maximum subarrays in about 52 seconds.

The rest of the paper is organized as follows.  In Section 2 we discuss previous results. In Section 3 we describe in more details the linear time divide and conquer solution in~\cite{bengtsson2007ranking}, for the maximum subarray problem (\textit{k} = 1). We then show how to extend this approach to find k-maximum subarrays in Section 4, and continue on to give our main, $O(n \sqrt{k})$ time divide-and-conquer algorithm, in Section 5. Section 5 also presents our $O(\sqrt{k} \log^{3} k)$ time solution for finding the 
$k$ largest sums of $X+Y$. We report on implementation, experimental results   and comparison with previous results in Section 6. 

\section{Previous Work}

Bengtsson, and Chen~\cite{bengtsson2006efficient} provided a complex, $O(min\{k + n \log^{2} n, n \sqrt{k} \})$ time algorithm to solve the \textit{k}-maximum subarray problem. Their main algorithm, for general $k$, has five phases. First, the problem is reduced to finding the top \textit{k} maximum values over all the "good" elements in some matrix of size $n \times n$. In the second phase, repeated constraint searches are performed, which decrease the number of candidate elements to $O(\min \{kn,n^2 \})$. In third phase, a range reduction procedure is performed to reduce the number of candidates further to $\theta(k)$. In the fourth phase, a worst-case linear-time selection algorithm is used on the remaining candidates, resulting in an element \textit{x}, that is the $k$-th largest sum. The final phase involves finding the "good" elements with values not less than \textit{x}. The $O(n \sqrt{k})$ part of the running time comes from a divide-and-conquer solution, and is useful for small values of $k$. The combine phase of the divide and conquer algorithm uses the $O(\sqrt{k})$ time algorithm from~\cite{frederickson1984generalized} to find the $k^{th}$ largest element in a sorted matrix, which is fairly difficlut to understand and tedious to implement. A trivial lower bound for this problem is $O(n + k)$. \\
In the same year (2006), Bae and Takaoka~\cite{bae2006improved} provided an $O((n+k) \log k)$ solution that reports the $k$ maximum subarrays in sorted order. \\
Still in 2006, Cheng, Chen, Tien, and Chao~\cite{cheng2006improved} provided an algorithm with running time $O(n + k \log(\min\{n, k\}))$. 
The authors adapted an iterative strategy where the table of possible subarray sums is build partially after every iteration and the algorithm terminates in $O(\log n)$ iterations, which yields a time complexity of $O(n + k \log(\min\{n, k\}))$. \\
Finally, in 2007, Bengtsson and Chen~\cite{bengtsson2007ranking}  provided a solution that takes time $O(n + k \log n)$ in the worst case to compute
and rank all $k$ subsequences. They also proved that their algorithm is optimal for $k = O(n)$, by providing a matching lower bound.
Their approach is different from the previous ones. In particular, although only briefly described, their solution provides an O($n$) time algorithm for the maximum subarray ($k$=1) problem. They give a tree based algorithm that uses a full binary tree, augmented with information about prefix sums, suffix sums, sums, and ranking among subsequences with respect to their sums. There are two phases of this algorithm. In the first phase, initial information (prefix sum, suffix sum, sum, largest elements) is computed and stored in the tree. The tree is constructed in a bottom-up fashion. The algorithm is based on the well known observation that the maximum sum can be obtained from the left branch or the right branch, or from a subsequence spanning over the left and right branches (subarrays). The second phase is the query phase which uses a binary heap to compute the \textit{k}-maximum subarrays. A special property of this algorithm is that if \textit{l} largest sums are already computed then the $(l+1)$-th largest sum can be found in $O(\log n)$ time. \\
Frederickson and Johnson~\cite{frederickson1984generalized} provided an efficient algorithm to find the $k^{th}$ maximum element of a matrix with sorted rows and columns. When the sorted matrix has $k$ rows and $k$ columns, their algorithm finds the $k^{th}$ largest element in O($\sqrt{k}$) time. 
It can then be used to find and report the $k$ largest values in the matrix in an additional $O(k)$ time. This corresponds to finding and reporting the $k$ largest values of $X+Y$.  
The algorithm is not simple, and is tedious to implement. 

\section{Linear time divide-and-conquer maximum subarray}
In this section we give a detailed description of a simple, linear time divide and conquer algorithm to find the maximum subarray (\textit{k} = 1), by placing the algorithm in~\cite{bengtsson2007ranking}
in a standard divide-and-conquer framework. 


\begin{figure}[t]
  \centering
  \includegraphics[scale=.26]{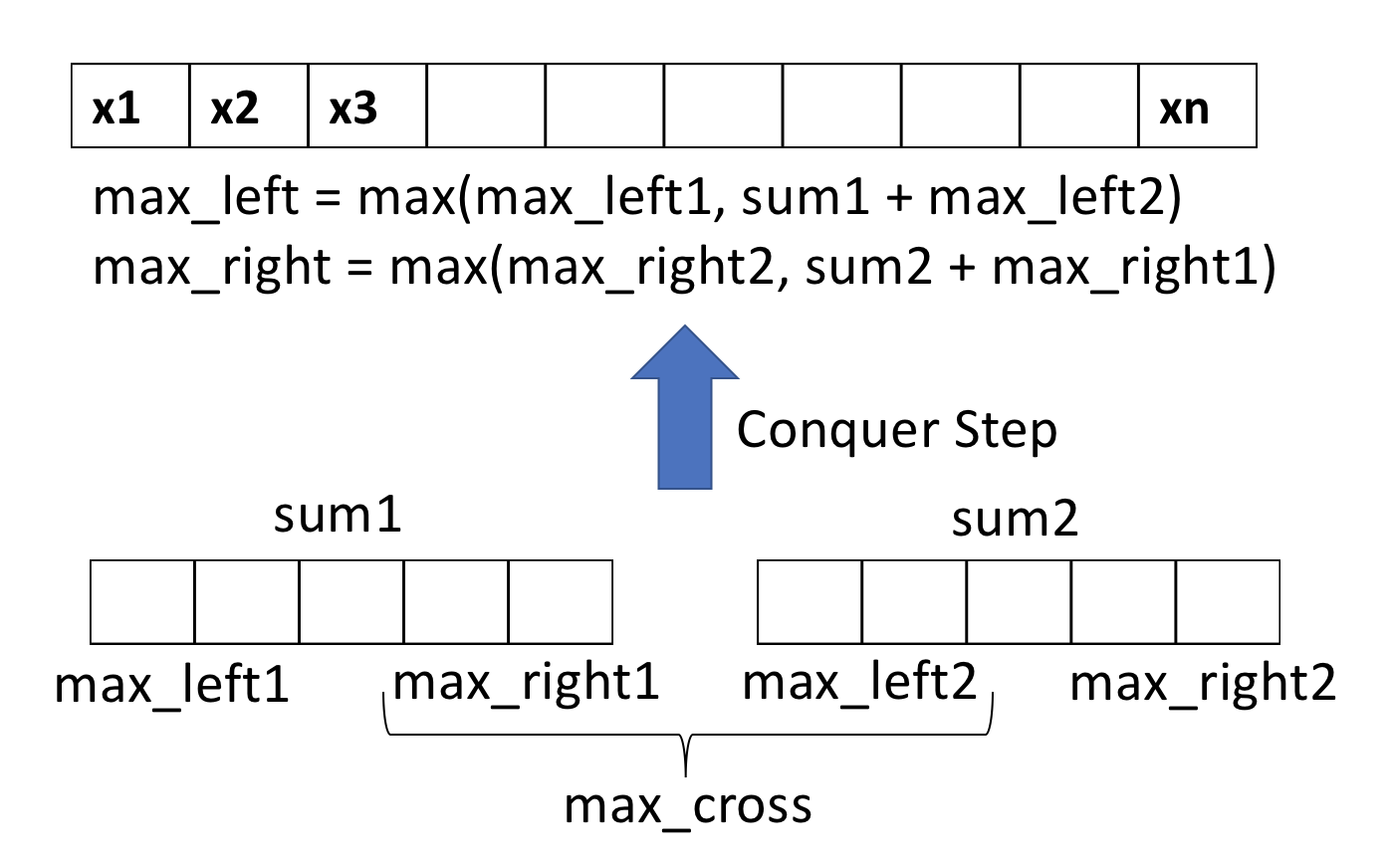}
  \caption{Illustration of combine phase}
  \label{fig1}
\end{figure}

Given an array \textit{A} of \textit{n} real numbers, the maximum subarray problem is to find a contiguous subarray whose sum of elements is maximum over all possible subarrays, including \textit{A} itself. The divide and conquer algorithm divides \textit{A}  into two subarrays of equal size, makes two recursive calls, and then proceeds with the combine step while keeping track of the maximum subarray sum found in the process. 

In the combine phase, at an internal node, we have two subarrays, $A_{1}$ (from left child) and $A_{2}$ (from right child). 
We define the following variables which are used to find the maximum subarray (see also 
Figure~\ref{fig1}): 
\\

\begin{tabular}{ |l|l| }
\hline
$max\_left \leftarrow -\inf$  &  maximum subarray starting from left most index \\ 
$max\_right\leftarrow -\inf$ & maximum subarray starting from right most index\\
$sum\leftarrow 0$ & sum of all elements in array\\
$max\_cross\leftarrow -\inf$ & maximum crossing subarray \\
$max\_sub\leftarrow -\inf$ & maximum subarray  \\
\hline
\end{tabular}
\\

The idea is to make the combine phase run in $O(1)$ time instead of the $O(n)$ time, as described
in~\cite{cormen2009introduction}.
For that, the values (and corresponding array indexes) of $max\_left$, $max\_right$, and $sum$ must
also be passed up from the recursive calls. The $sum$ value at a given node can be
found by adding up the sums from the children. The value $max\_left$ is either the $max\_left$
from the left child or the $sum$ value from the left plus the $max\_left$ value from the right child. Similarly, the value $max\_right$ is either the $max\_right$
from the right child or the $sum$ value from the right plus the $max\_right$ value from the left child. 
The following divide and conquer algorithm, $Maximum\_Subarray$, takes in the input an array $A$ of size \textit{n} and two integers, \textit{low} and \textit{high}, which correspond to the start index and end index of subarray $A[low \dots high]$, and finds and returns the maximum subarray of $A[low,high]$. 
\begin{algorithm}\textbf{Maximum$\_$Subarray (A, low, high)}
\begin{enumerate}
\item if ($low$ == $high$)
\item \qquad	$max\_left$ = A[low];
\item	\qquad $max\_right$ = A[low];
\item \qquad	$sum$ = A[low];
\item \qquad	$max\_sub$ = A[low];
\item \qquad	return ($max\_left, max\_right, sum, max\_sub$)
\item mid = $\lfloor$ $\frac{low + high}{2}$ $\rfloor$
\item $(max\_left1, max\_right1, sum1, max\_sub1)$=\textbf{Maximum$\_$Subarray$(A, low, mid)$}
\item $(max\_left2, max\_right2, sum2, max\_sub2)$=\textbf{Maximum$\_$Subarray$(A, mid+1, high)$}
\item $max\_left$ = max($max\_left1, sum1 + max\_left2$);
\item $max\_right$ = max($max\_right2, sum2 + max\_right1$)
\item $sum$ = $sum1 + sum2$
\item $max\_cross$ = $max\_right1 + max\_left2 $
\item $max\_sub$ = max($max\_cross, max\_sub1, max\_sub2$)
\item return ($max\_left, max\_right, sum, max\_sub$)
\end{enumerate}

\end{algorithm} 

In above algorithm, steps 1-7 take O(1) time. Steps 8-9 correspond to the recursive calls. Steps 10-15 take O(1) time. Therefore, the time taken by Algorithm 1 is: $ T(n) = 2T(n/2) + O(1) = O(n)$.

In the next section, we extend this algorithm to find the k-maximum subarrays.
\section{k-maximum subarrays by divide and conquer}

Given an array \textit{A} of $n$ numbers and an integer \textit{k}, where $1 \leq k \leq n(n+1)/2$, the $k$-maximum subarrays problem is to find the $k$ contiguous subarrays with the largest sums. 
In this section we address the $k$-maximum subarrays problem and provide a warm-up 
divide and conquer solution. 

We extend the approach in Section 3, used to find the maximum subarray. We recursively divide the array into two subarrays with equal elements until we reach a base case (of size $\sqrt{k}$), and then perform the combine step. The main difference, and implied difficulty, is that recursive calls return information about $k$ largest subarrays from corresponding subproblems, including $k$ largest sums from the left and from
the right, and we are finding \textit{k} largest subarray values in the combine step. 
A detailed description of the generic algorithm is given in~\cite{bengtsson2006efficient}.
Our main goal is to simplify the combine phase. 

In the remaining of the paper, notations like $max\_left$, $max\_right$, $max\_cross$, and 
$max\_sub$ refer to
arrays of size $k$, holding the corresponding $k$ largest sum values. Except for
$max\_cross$, these arrays are sorted in non-increasing order. 

Consider the left and right subarrays, $A_{l}$ and $A_{r}$, of some internal node $v$ in the recursion tree. The $k$ largest sums at $v$ are among the $k$ largest sums from $A_l$, the $k$ largest sums from $A_r$, and the $k$ largest sums of contiguous  subarrays that cross between $A_l$ and $A_r$ (we call these last sums {\it crossing sums}). The difficult part is to efficiently compute the $k$ crossing sums and the various $k$ largest sums that need to be passed up to the parent node.
 
In this section we provide a solution for the combine step that is simple yet efficient, easy to implement, and sets up the stage for our better solution in the following section. 

The function \textbf{MERGE} used in the following algorithms is similar to the one in the {\it merge-sort} sorting algorithm, except that we stop after finding the largest $k$ values, and takes O(k) time. By a slight abuse, we allow the MERGE function to work with a constant number of arrays in the input, rather than just two arrays. If there are more than two arrays passed to the MERGE function we perform pairwise merge to find the $k$ largest numbers.
Similarly, function \textbf{SELECT}, whenever mentioned, is the standard linear time selection function~\cite{blum1973time}, that finds the $k$-th (and thus $k$) largest number(s) of a given set of 
($O(k)$ in our case) numbers.

The function \textbf{MAX$\_$SUM $(a, A)$} used below takes in the input an array \textit{A} of size $k$ and an integer \textit{a} and adds $a$ to each entry of \textit{A}. It is used to add the value of the
sum of elements of the left (or right) child of $v$ to the $k$ largest sum prefix (suffix) values of the right (resp., left) child of $v$.  

The following procedure takes as input two arrays, \textit{A} and \textit{B}, each of size $k$, sorted in non-increasing order, and outputs the \textit{k}-maximum sums of the pairwise addition of $A$ and $B$. For our purpose, $A$ would contain the $k$ largest sums of $A_l$ for subarrays starting at the rightmost entry of $A_l$, while $B$ would contain the $k$ largest sums of $A_r$ for subarrays starting at the leftmost entry of $A_r$. We use a priority queue $Q$ implemented as a binary heap to store pairwise sums, as they are generated. An AVL tree $T$ is also used, to avoid placing duplicate pairs $(i,j)$ in $Q$.
\begin{algorithm} \textbf{MAX$\_$SUM$\_$CROSS (A, B)}
\begin{enumerate}
\item k = sizeof(A)
\item Q $\leftarrow$ null					\hfill//Max Priority Queue
\item M[k] $\leftarrow$ null; 				\hfill	//Output Array
\item T $\leftarrow$ null;					\hfill//AVL Tree
\item m $\leftarrow$ 0;
\item add $(0, 0)$ to Q with priority $A[0] + B[0]$
\item store $(0, 0$) in T
\item while k $> 0$:
\item \qquad	$(i, j)$ = pop Q 
\item	\qquad M[m] = A[i] + B[j]
\item \qquad	m = m + 1; k = k - 1
\item \qquad	if $( i < k$ and $(i+1, j) \not \in$ T) 
\item	 \qquad \qquad	store (i+1, j) in T
\item	 \qquad \qquad	add $( i + 1, j )$ to Q with priority $(A[i+1] + B[j])$
\item	\qquad if $( j < k$ and $(i, j+1) \not \in$ T) 
\item	\qquad \qquad	store $(i, j+1)$ in T
\item	\qquad \qquad	add $( i, j + 1 )$ to Q with priority $(A[i] + B[j+1]) $
\item return M

\end{enumerate}
 \end{algorithm}

\noindent \textbf{Time Complexity of Algorithm MAX$\_$SUM$\_$CROSS}:
Lines 12, 15 take $O(\log k)$ time for searching T, lines 13, 16 take $O(\log k)$ time to store indices in T, lines 9, 14, 17 take $O(\log k)$ to add or remove an element in the priority queue,
and the while loop in line 8 runs $k$ times. Therefore, the time complexity for 
algorithm MAX$\_$SUM$\_$CROSS  is O($k \log k$).

The \textbf{Max-k} algorithm below computes the \textit{k}-maximum sums (subarrays) of the given array \textit{A}. The values \textit{low} and \textit{high} correspond to the start index and end index of the subarray $A[low \ldots high]$. 
\begin{algorithm}\textbf{Max-k (A, low, high)}
\begin{enumerate}
\item if ($low$ + $\sqrt{k}$ $\geq$ $high$) then find $max\_left, max\_right, sum, max\_sub$ by brute force and return ($max\_left, max\_right, sum, max\_sub$)
\item mid = $\lfloor$ $\frac{low + high}{2}$ $\rfloor$
\item $(max\_left1, max\_right1, sum1, max\_sub1)$ = \textbf{Max-k}  $(A, low, mid)$
\item $(max\_left2, max\_right2, sum2, max\_sub2)$ = \textbf{Max-k} $(A, mid+1, high)$
\item $max\_left$ = \textbf{MERGE$(max\_left1,$ MAX$\_$SUM $(sum1, max\_left2))$;
\item $max\_right$ = MERGE$(max\_right2,$  MAX$\_$SUM $(sum2, max\_right1))$
\item $sum$ = $sum1 + sum2$
\item $max\_cross$ = MAX$\_$SUM$\_$CROSS$(max\_right1, max\_left2 )$
\item $max\_sub$ = MERGE$(max\_cross, max\_sub1,$ $max\_sub2)$}
\item return ($max\_left, max\_right, sum, max\_sub$)
\end{enumerate}
 \end{algorithm}

\noindent \textbf{Running time of Max-k:} 
It can be easily seen that the running time of algorithm \textbf{Max-k} is described by the recurrence: 

$T(n,k) = 2T(n/2,k) + O(k \log k ),$ with $T(\sqrt{k},k)=O(k)$. 

Using substitution method, we have $T(n , k) = 2^{i} T(n/2^{i} , k) + O(\sum_{j = 0}^{i-1} 2^j k \log k)$. \quad Letting $(n / 2^{i})^{2} = k$ results in $2^{i} = n/\sqrt{k}$.

$T(n, k) = (n / \sqrt{k}) T(\sqrt{k}, k) + O(n \sqrt{k} \log k)$. Since $T(\sqrt{k}, k) = O(k)$ the overall time complexity is $ O(n \sqrt{k} \log k)$. The algorithm is an $O(\log k)$ factor slower than the one in~\cite{bengtsson2006efficient}, while being very simple to describe and implement.  

In the next section we provide a simple, $O(k)$ time prune-and-search algorithm for the $MAX\_SUM\_CROSS$ procedure, which improves the overall time complexity of algorithm \textbf{Max-k} to
$O(n \sqrt{k})$. 

\section{An improved algorithm for $k$-maximum subarrays}

In this section, we improve the results in the previous section by providing an $O(k)$ time divide-and-conquer solution for the combine phase. 
To this end, we first find the $k^{th}$ largest element \textit{x} of the pairwise sum $A + B$ ~\cite{frederickson1982complexity,frederickson1984generalized}, and then scan \textit{A} and \textit{B} for elements in $A + B$ greater than or equal to \textit{x}. If this output would be sorted then it will again lead to an O($k \log k$) running time. However, as explained later in this section, there is no need to sort these elements, that correspond to the values of the crossing sums. 

Frederickson and Johnson~\cite{frederickson1984generalized} provided an algorithm that can find the $k^{th}$ maximum element of a matrix consisting of \textit{k} rows and \textit{k} columns, each sorted in nonincreasing order, in O($\sqrt{k}$) time. 
 Given a sorted matrix \textit{M}, the algorithm extracts a set $S$ of submatrices of different shapes which guarantee to contain all elements greater than or equal to the $k^{th}$ largest element of \textit{M}. Note that they also contain elements which are less than the $k^{th}$ largest element. After extracting the set of submatrices, it forms a new matrix with the help of  dummy matrices (matrices where all entries are $-\infty$). The new matrix is also sorted. The submatrices are referred to as \textit{cells}, and for each cell \textit{C}, min(\textit{C}) and max(\textit{C}) represent the smallest and largest elements in this cell. Initially, there is a single cell which is the matrix formed from dummy matrices and the set $S$. After each iteration, a cell is divided into four subcells. From all the subcells formed the algorithm computes some values that allow to discard a few cells guaranteed not to contain the $k^{th}$ largest element.
The algorithm is not simple to follow, and tedious to implement. 
Also the construction of sorted matrix from set of sorted submatrices and dummy matrices is not easy to understand and implement.



Intuitively, for our problem, the rows and columns of the matrix are generated by the sums in $A+B$, where $A$ and $B$ are sorted arrays of size $k$ each. In row $i$, $A[i]$ is summed over the entries in array $B$. Similarly, in column $j$, $B[j]$ is summed over the entries in array $A$. 
The matrix does not have to be explicitly stored as 
the matrix entries can be generated as needed from the values in $A$ and $B$.
Thus, using the algorithm in~\cite{frederickson1984generalized} one can compute the $k^{th}$ maximum element $x$ of $A+B$
in $O(\sqrt{k})$ time. Retrieving the elements of $A+B$ that are greater than (or equal) to $x$ takes an additional $O(k)$ time. 
This makes the algorithm \textbf{MAX$\_$SUM$\_$CROSS} in previous section run in $O(k)$ time. 
Since the $k$ largest crossing sum values are no longer sorted, we replace the \textbf{MERGE}
call in line 9 of algorithm \textbf{Max-k} with a \textbf{SELECT} call. 
The algorithm, as presented above, has been described in~\cite{bengtsson2006efficient}. 

The only place where the $k$ largest crossing sum values are used at an internal node $u$ 
of the recursion tree is in the calculation of the $k$ largest sum values at $u$, given 
the $k$ largest sum values from the left and the right children of $u$. Let $v$ be the parent node of $u$.
Node $u$ needs to pass up to $v$ the $k$ largest sum values of subarrays that start at the
leftmost entry, and the $k$ largest sum values of subarrays that start at the
rightmost entry ($max\_left$ and $max\_right$ arrays at $u$) and these subarrays are either distinct from the crossing subarrays at $u$ or
computed independently of those subarrays by function \textbf{MAX\_SUM}. See Figure~\ref{fig1}.

The following lemma is implicitly used in~\cite{bengtsson2006efficient}. 

\begin{lemma}
The $k$ largest crossing sum values do not need to be sorted for algorithm \textbf{Max-k} to
correctly report the $k$ largest sum values of $A$.
\end{lemma}



Then, the running time of the \textbf{Max-k} algorithm is now described by the recurrence
$T(n, k) = 2T(n/2, k) + O(k),$ with $T(\sqrt{k},k)=O(k)$
As described in~\cite{bengtsson2006efficient}, using the substitution method, we have 
$T(n , k) = 2^{i} T(n/2^{i} , k) + O(\sum_{j = 0}^{i-1} k 2^{j})$
Letting $(n / 2^{i})^{2} = k$ results in $2^{i} = n/\sqrt{k}$.

$T(n, k) = (n / \sqrt{k}) T(\sqrt{k}, k) + O(n \sqrt{k})$.
Thus,  $T(n)= O(n \sqrt{k})$.

As mentioned earlier, the algorithm for finding the $k$-th largest entry in $A+B$, as presented in~\cite{frederickson1984generalized}, is complex and tedious to implement. 
In what follows, we provide a simple algorithm to find the $k$-th largest element in $A+B$, which takes O($\sqrt{k} \log^{3} k$) time and is easy to implement. Moreover, unlike the
algorithm in~\cite{frederickson1984generalized}, our algorithm is a simple prune-and-search procedure. Also, unlike in~\cite{frederickson1984generalized}, our algorithm implicitly finds the \textit{k} largest elements of A + B in the process.
The total time needed to report all \textit{k} largest elements in $A+B$ is then O($k + \sqrt{k} \log^3 k$) which is still $O(k)$, and thus the final time complexity of algorithm \textbf{Max-k} remains $O(n \sqrt{k})$.

\begin{figure}[t]
\centering
\includegraphics[scale=0.14]{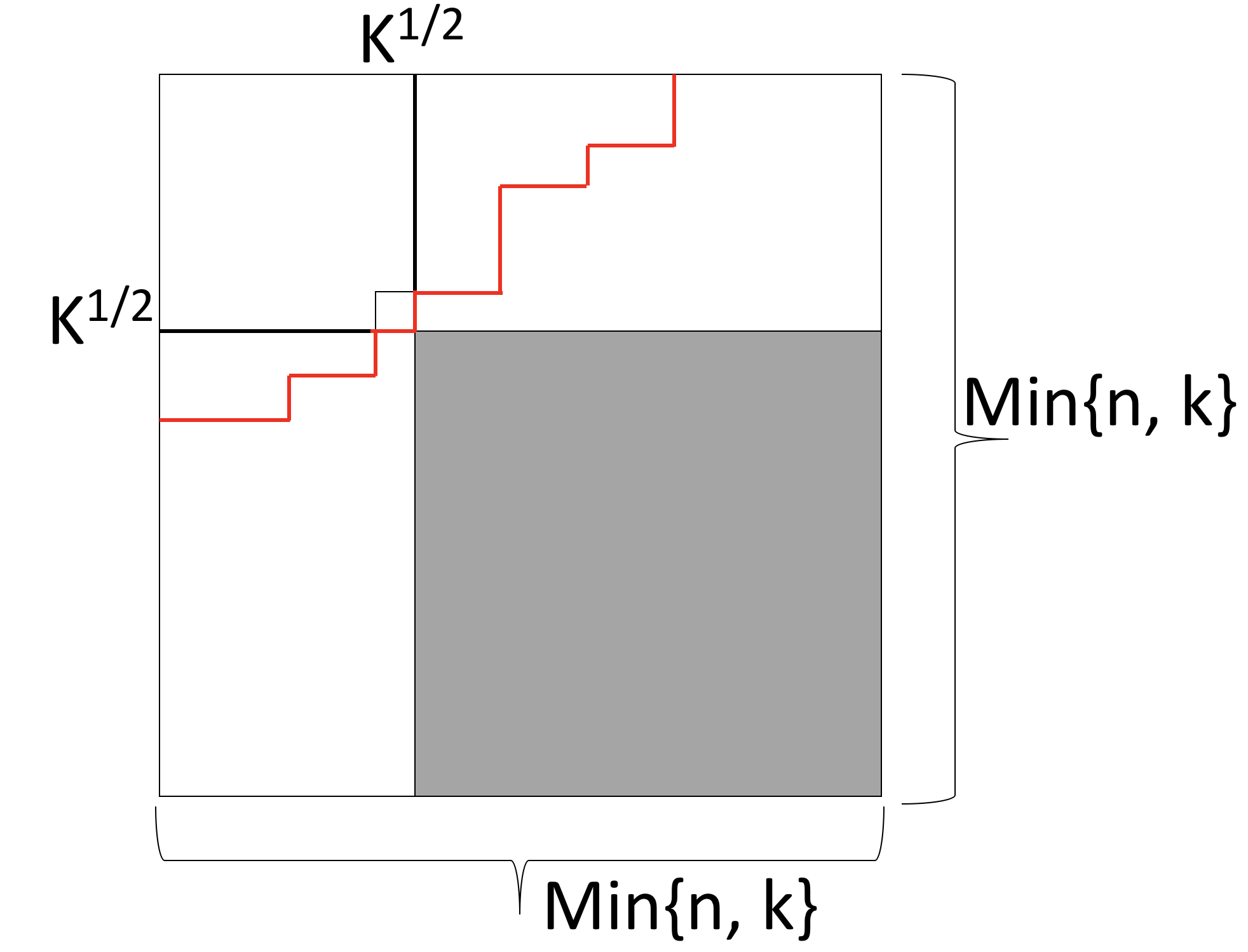}
\caption{Initial staircase for $x=M[\sqrt{k},\sqrt{k}]$; the shaded region of $M$ is irrelevant.}
\label{fig2}
\end{figure}

Let A and B be arrays of size \textit{n} and let \textit{k} be an interger such that $k \leq n^{2}$. We now show how to find the $k^{th}$-largest element of A + B and an implicit representation of the k largest elements of A + B in sublinear, O($\sqrt{k} \log^{3} k$) time.

Consider a matrix \textit{M} with $n$ rows and $n$ columns, 
such that the element in matrix entry $M[i, j]$ is the sum $A[i] + B[j]$. We have $\sqrt{k} \leq n$.
It is easy to observe that all rows and columns in matrix \textit{M} are sorted and the $k$ largest values will not lie in the submatrix $M[\sqrt{k}, \sqrt{k}$; $\min\{n,k\}$, $\min\{n,k\}]$, which is shaded in Figure~\ref{fig2}. 
Without loss of generality, assume that $k \le n$.
We call $M$ a \textit{sorted matrix}. Note that for the k-maximum subarray problem, matrix M is of size $k\times k $.

Matrix \textit{M} is only considered for better understanding of the algorithms presented in this section, but there is no need to store it explicitly. Instead, its entries are computed only as needed. The only information required is the start and end index of each row $\in [1, \sqrt{k}]$ and each column $\in [1, \sqrt{k}]$, that define the "active" entries of $M$ at a given step. Whenever a matrix is passed in a function, we are passing arrays, storing the start and end indexes of these rows and columns. 

A \textit{staircase} matrix $M_S$ is a subset of adjacent rows and columns of $M$, where each row and each column are described by a start and an end index. For a row (column) $i+1$, its start index is no larger than the start index of row (column) $i$ and its end index is no larger than the 
end index of row (column) $i$. 

Let max$\_$cross be an array of size $k$. Initially, all entries of max$\_$cross are set to minus infinity. Let $p \le \sqrt{k}$ and $m, r \le k$ be positive integers. \\
The notation $M[i][0:j]$ denotes the entries in row $i$ of the matrix $M$,
columns 0 to $j$. The notation $M[0:i][j]$ denotes the entries in column $j$ of matrix $M$, 
rows 0 to $i$. 

Given matrix \textit{M} and a pair ($m,r$) such that $A[m] + B[r] \geq M[\sqrt{k}, \sqrt{k}]$, the following algorithm will find and return: \\
(i) a staircase matrix of $M$ where all elements are greater than or equal to $M[m,r]$ \\
(ii) the total number $T$ of elements in the staircase matrix.
\begin{algorithm}\textbf{STAIRCASE(M, m, r, p)}
\begin{enumerate}
\item $x = M[m, r]$ 
\item for each row $i$ $\in [1, p]$ of $M$, use binary search to find the maximum index $\alpha_{i}$ such that $M[i][0: \alpha_{i}] \geq x$. 
\item for each column $j$ $\in [1, p]$ of $M$, use binary search to find the maximum index $\beta_{j}$ such that $M[0:\beta_{j}][j] \geq x$. 
\item Let $M_S$ be the (implicitly defined, staircase) submatrix of $M$ formed by elements larger or equal than $x$ found in step 2 and step 3.
\item Let T be the total number of elements in $M_S$
\item return $M_S$, T
\end{enumerate}
\end{algorithm}

In algorithm \textbf{STAIRCASE}, binary search on each row or column requires O($\log k$) time. There are \textit{p} rows and \textit{p} columns, and $p \leq \sqrt{k}$. The staircase matrix $M_S$ is defined implicitly, by start-end pairs for rows and columns. Therefore the total time of algorithm \textbf{STAIRCASE} is  $O(p \log k$). 

Notice that we need to pay attention to not double count the entries in $M[0:p][0:p]$, which can be easily done in $O(p)$ time.


\begin{figure}
\centering
\includegraphics[scale=0.14]{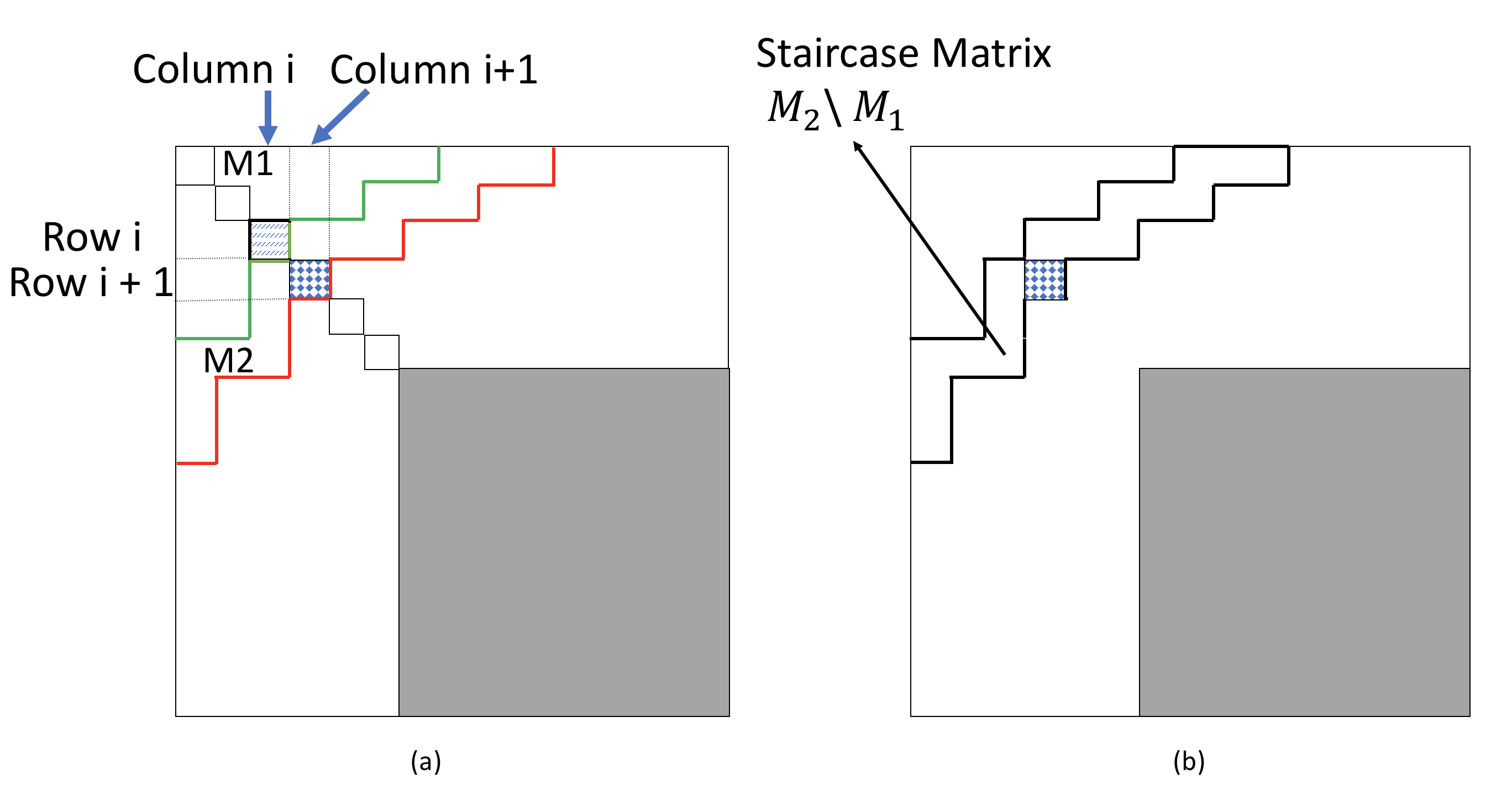}
\caption{(a) Two staircase matrices $M_{1}$ and $M_{2}$. Matrix $M_{1}$ has total number of elements less than or equal to $k$ while matrix $M_{2}$ has more than $k$ elements.(b) Illustrating $M_{2} \setminus M_{1}$}
\label{fig3}
\end{figure}

The idea behind the following algorithm is to find two consecutive diagonal index \textit{i} and \textit{i+1} of matrix \textit{M} such that the staircase matrix computed by $STAIRCASE(M, i, i, i)$ contain only elements which are greater than or equal to the $k^{th}$ largest element of matrix \textit{M}, while elements which are not part of the staircase matrix $STAIRCASE(M, i+1, i+1, i+1)$ are guaranteed to be less than the $k^{th}$ largest element of \textit{M}. It is easy to notice that the $k^{th}$ largest element lies in the subtraction of the matrices $STAIRCASE(M, i+1, i+1, i+1)$ and  $STAIRCASE(M, i, i, i)$.

The following algorithm takes as input a sorted matrix \textit{M} and computes a staircase submatrix of $M$ containing the $k$ largest entries in $M$. As we will see, it does that in $O(p\log^3 k)$ time, using an implicit representation of submatrices of $M$. The $k$ largest entries can then be reported in an additional $O(k)$ time. 
\begin{algorithm}\textbf{MAX$\_$SUM$\_$CROSS-1(M)}
\begin{enumerate}
\item Use binary search on $1,2,\ldots,\sqrt{k}$ to find index $i$ such that the total number of elements returned by STAIRCASE($M, i, i, i$) is at most $k$ and the total number of elements returned by STAIRCASE($M, i+1, i+1, i+1$) is greater than $k$ (This binary search on the diagonal of matrix M is illustrated in Figure~\ref{fig3}-a). 
\begin{enumerate}
\item 	Let $M_{1}, T_{1}$ = STAIRCASE($M, i, i, i$). 
\item 	Let $M_{2}, T_{2}$ = STAIRCASE($M, i+1, i+1, i+1$)
\end{enumerate}

\item if $T_{1} = k$
\item \qquad	return $M_{1}$
\item totalElementsLeft = $k - T_{1}$ 	\hfill$//elements$ to be $found$ in $M_{2}\setminus M_{1}$
\item $M_S = M_{2} \setminus M_{1}$
\item $M_{final}$ = FIND$\_$INDEX($M_S$, totalElementsLeft, i+1)
\item return $M_{final}$
\end{enumerate}
 \end{algorithm}

In algorithm \textbf{MAX$\_$SUM$\_$CROSS-1}, $M_{2} \setminus M_{1}$ corresponds to the staircase matrix formed by deleting elements of matrix $M_{1}$ from $M_{2}$. Step 1 requires 
O($p \log^{2} k$) and finds the tuples $(M_1,T_1)$ and $(M_2,T_2)$. 
In step 5, we store $O(p)$ indexed pairs into $M$, which takes $O(p)$ time. 
Let $\Gamma$ be the running time for algorithm \textbf{FIND$\_$INDEX} (step 6).
The total time taken by algorithm \textbf{MAX$\_$SUM$\_$CROSS-1} is then O(max$\{p \log^{2} k, \Gamma  \}$).

Let $M_S$ be the staircase matrix which corresponds to $M_{2} \setminus M_{1}$. $M_S$ is 
implicitly defined and stored. For row \textit{i} of $M_S$, let the lowest index be $i_l$, and the highest index be $i_h$. The median value of each row can be found in constant time, at entry $(i_l+i_h)/2)$. 
A similar notation is used to denote the lowest and the highest entries of a column of $M_S$. 

Algorithm \textbf{FIND$\_$INDEX} below takes as input $M_S$, an integer which stores the rank of the element we need to find in $M_S$, and an index \textit{p} useful for computing staircase matrices, and returns the $k^{th}$ largest elements of matrix \textit{M} in an implicit representation.  
\begin{algorithm}\textbf{FIND$\_$INDEX($M_S$, totalElementsLeft, $p$)}
\begin{enumerate}
\item Find the median value in each row 1 to $p$ and in each column 1 to $p$ of $M_S$ and place them into an array $X$. 
\item Sort $X$ in non-increasing order. For element $x_{i}$ at $i$-th position in array X, let $\alpha_{i}$ be the total number of elements of $M_S$ greater than or equal to $x_{i}$ and let $\beta_{i}$ be the total number of elements of $M_S$ strictly greater than $x_{i}$. 
\item Use binary search on $X$ together with the STAIRCASE function to find the maximum index $i$ and the minimum index $j$ in array $X$ such that $totalElementsLeft - \alpha_{i} \geq 0$ and $totalElementsLeft - \alpha_{j} < 0$. Find corresponding $\beta_{i}$ and $\beta_{j}$. Notice that $j = i + 1$. When searching, the last argument passed to the STAIRCASE function is $p$, while $m$ and $r$ correspond to the row and column in $M_S$ for the current search value in $X$.
\item if $\exists$  $i, j$ in step 3, 
\begin{enumerate}
	\item if $totalElementsLeft - \alpha_{i} = 0$
	\begin{enumerate}
		\item Return staircase matrix of $M$ with smallest element as $x_{i}$
	\end{enumerate}
	\item else if $totalElementsLeft - \beta_{j} = 0$
	\begin{enumerate}
		\item Return staircase matrix of $M$ with all elements greater than $x_{j}$
	\end{enumerate}
	\item else if $totalElementsLeft - \beta_{j} < 0$
	\begin{enumerate}
		\item Let $M^{'}$ be the staircase matrix obtained by removing all elements greater or equal to $x_{i}$ and less than or equal to $x_{j}$ from $M$
		\item totalElementsLeft = totalElementsLeft - $\alpha_{i}$
		\item FIND$\_$INDEX($M^{'}$, totalElementsLeft, $p$)
	\end{enumerate}
	\item else if $totalElementsLeft - \beta_{j} > 0$
	\begin{enumerate}
		\item totalElementsLeft = totalElementsLeft  - $\beta_{j} $
		\item Return staircase matrix of $M$ with all element greater than $x_{j}$ and $totalElementsLeft$ number of element equal to $x_{j}$	
	\end{enumerate}
\end{enumerate}
\item else if $\exists$ $j$ and $\not \exists$ $i$
\begin{enumerate}
	\item if $totalElementsLeft - \beta_{j} = 0$
	\begin{enumerate}
		\item Return staircase matrix of $M$ with all element greater than $x_{j}$
		\end{enumerate}
	\item else if $totalElementsLeft - \beta_{j} < 0$
	\begin{enumerate}
		\item Let $M^{'}$ be the staircase matrix obtained by removing all elements less than or equal to $x_{j}$  from $M_S$
		\item FIND$\_$INDEX($M^{'}$, totalElementsLeft, $p$)
		\end{enumerate}
	\item else if $totalElementsLeft - \beta_{j} > 0$
	\begin{enumerate}
		\item totalElementsLeft = totalElementsLeft  - $\beta_{j} $
		\item Return staircase matrix of $M$ with all element greater than $x_{j}$ and $totalElementsLeft$ number of element equal to $x_{j}$
		\end{enumerate}
		
\end{enumerate}
\item else if $\exists$ $i$ and $\not \exists$ $j$
\begin{enumerate}

	\item if $totalElementsLeft - \alpha_{i} = 0$
	\begin{enumerate}
		\item Return staircase matrix of $M$ with smallest element as $x_{i}$
		\end{enumerate}
	\item else 
	\begin{enumerate}
		\item Let $M^{'}$ be the staircase matrix obtained by removing all elements greater than or equal to $x_{i}$ from $M_S$
 		\item totalElementsLeft = totalElementsLeft - $\alpha_{i}$
 		\item FIND$\_$INDEX($M^{'}$, totalElementsLeft, $p$)
 		\end{enumerate}
 
\end{enumerate}
 
\end{enumerate}
 \end{algorithm}




We now analyze the running time of algorithm \textbf{FIND$\_$INDEX}.
Step 1 requires O($p$) time, Step 2 requires O($p \log k$) time, and Step 3 requires 
O($p \log^{2} k$) time. Computing the staircase matrix and the total elements in steps 4, 5, and 6 requires O($p \log k$) time. In steps 4, 5, and 6, half of elements are removed before function \textbf{FIND$\_$INDEX} is called recursively. With $p=O(\sqrt{k})$, the time complexity is then described by
$T(k, p) = T(k/2, p) + p \log^{2}k,$ with $T(1,p)=O(1)$

Using substitution method, we have 
$T(k, p) = T(k / 2^{i}, p) + O(\sum_{j = 0}^{i-1}p \log^{2} (k/2^{j}))$

Letting $k / 2^{i} = 1$ results in $i = \log k$ and

$T(k, p) = T(1, p) + O(\sum_{j = 0}^{\log k-1} p \log^{2} (k/2^{j}))$

which solves for $T(k,p)=O(p \log^{3} k$).  With $\Gamma = O(p \log^{3} k$), algorithm \textbf{MAX$\_$SUM$\_$CROSS-1} 
thus takes O($p \log^{3} k$) time where $p = O(\sqrt{k})$.
We summarize our result below.

\begin{theorem}
Algorithm \textbf{MAX$\_$SUM$\_$CROSS-1} finds the $k$ largest elements in $A+B$ (and thus the $k$ 
largest crossing sums) in O($k \log^{3} k$) time. The $k$ sums are implicitly represented and can be report with an additional $O(k)$ time.
\end{theorem}

Recall that there is no need to sort the elements in the $max\_cross$ since the arrays contributing in the combine step are $max\_left$ and $max\_right$. Therefore, instead of using $MERGE$, we can use the $SELECT$ algorithm in line 9 of algorithm \textbf{Max-$k$} to find \textit{k} $max\_sub$ in O($k$) time. 


Plugging in the new procedure for finding crossing sums, \textbf{MAX$\_$SUM$\_$CROSS-1}, 
the running time of the divide and conquer algorithm 
\textbf{Max-$k$} is now described by the recurrence 
$T(n, k) = 2T(n/2, k) + O(k), with T(\sqrt{k},k)=O(k)$

\noindent which solves for $T(n) = O(n \sqrt{k})$.

\begin{theorem}
Algorithm \textbf{MAX-$k$} finds the $k$ largest subarrays of an array $A$ of size $n$ in 
$O(n \sqrt{k})$ time, where $1 \le k \le n(n+1)/2$. 
\end{theorem}




\section{Implementation and Experiments}

We have implemented our algorithms and performed multiple experiments, reporting excellent results. For comparision, we also implemented the algorithm presented in~\cite{frederickson1984generalized} which is used in the combine phase in~\cite{bengtsson2006efficient}. The implementation is in JAVA on macOS High Sierra with 3.1 GHz intel i5 processor and 8 GB of RAM, while the data sets have been randomly generated.

For experimentation, we generated two random sorted arrays of size \textit{n} which consist of integer values and defined an integer \textit{k}. For comparision both \textit{n} and \textit{k} are power of 4 as assumed in~\cite{frederickson1984generalized}. Table 1 shows comparision results between the two algorithms. It is clear that our algorithm outperformed. The time complexity of our algorithm depends upon the value of $p \leq \sqrt{k}$.  At each step of our algorithm, we eliminate many elements which are not candidates for the $k^{th}$ largest element. For example, while computing a staircase matrix we have knowledge about each row and column (elements in submatrix $M[1, m; 1, r]$ are greater than or equal to $M[m, r]$). 
%

\begin{table}
\caption{Comparison between our algorithm and~\cite{frederickson1984generalized} ($k=n$).}
\centering
\begin{tabular}{| l || l || l |}
\hline
Size of input array & \multicolumn{2}{c|}{Average time for $10^{2}$ test cases 
(in milliseconds)}\\
\hline
 & Algorithm in~\cite{frederickson1984generalized} & Our Algorithm \\
 \hline 
 $4^{4}$ & 0.18 & 0.03 \\
 $4^{5}$ & 0.91 & 0.12 \\
 $4^{6}$ & 1.2 & 0.14 \\
 $4^{7}$ & 2.6 & 0.17\\
 \hline

\end{tabular}
\label{table:2}
\end{table}

\begin{table}
\caption{Average time taken to find the \textit{k}-maximum values of $A+B$ ($k=n$).}
\centering
\begin{tabular}{|p{3.0cm}||p{3.0cm} ||p{3.0cm} ||p{3.0cm} |}
\hline
Size of input array & \multicolumn{3}{c|}{Average Time (in milliseconds)}\\
\hline
	& Number of tests = $10^{2}$ & Number of tests = $10^{3}$ & Number of tests = $10^{4}$ \\
\hline
10 & 0.03 & 0.01 & 0.02\\
$10^{2}$ & 0.07 & 0.03 & 0.04\\
$10^{3}$ & 0.18 & 0.08 & 0.09\\
$10^{4}$ & 0.37 & 0.23 & 0.27\\
$10^{5}$ & 1.35 & 1.44 & 1.39\\
$10^{6}$ & 16.21 & 15.78 & 15.85\\
\hline

\end{tabular}
\label{table:2}
\end{table}

\begin{table}
\caption{Average time taken to find the \textit{k}-maximum subarrays when k = n.}
\centering
\begin{tabular}{|p{3.0cm}||p{3.0cm} ||p{3.0cm} ||p{3.0cm} |}
\hline
Size of input array & \multicolumn{3}{c|}{Average Time (in milliseconds)}\\
\hline
	& Number of tests = $10^{2}$ & Number of tests = $10^{3}$ & Number of tests = $10^{4}$ \\
\hline
10 & 0.08 & 0.03 & 0.01 \\
$10^{2}$ & 0.44 & 0.363 & 0.10\\
$10^{3}$ & 4.19 & 3.63 & 1.60 \\
$10^{4}$ & 91.28 & 98.56 & 93.43 \\
$10^{5}$ & 1729.78 & 1890.03 & 1780.13\\
$10^{6}$ & 52101.62 & - & -\\ 

\hline

\end{tabular}
\label{table:3}
\end{table}

\begin{table}
\caption{Average time taken to find the \textit{k}-maximum subarrays for small k.}
\centering
\begin{tabular}{|p{3.0cm}||p{1.4cm} ||p{1.4cm} ||p{1.4cm} ||p{1.4cm}||p{1.4cm} |}
\hline
Size of input array & \multicolumn{5}{c|}{Average Time (in milliseconds)}\\
\hline
	& k=5 & k=15 & k=25 & k=35 & k=45  \\
\hline
10 & 0.07 & 0.04 & 0.02 & 0.03 & 0.02\\
$10^{2}$ & 0.31 & 0.33 & 0.24 & 0.31 & 0.28\\
$10^{3}$ & 0.81 & 2.03 & 1.45 & 1.21 & 0.88\\
$10^{4}$ & 4.03 & 4.78 & 6.2 & 5.39 & 5.02\\
$10^{5}$ & 34.63 & 45.11 & 46.33 & 51.58 & 46.94\\
\hline

\end{tabular}
\label{table:4}
\end{table}

\begin{table}
\caption{Average time taken to find the \textit{k}-maximum subarrays.}
\centering
\begin{tabular}{|p{3.0cm}||p{1.3cm} ||p{1.3cm} ||p{1.3cm} ||p{1.3cm}||p{1.3cm} ||p{1.3cm} |}
\hline
Size of input array & \multicolumn{6}{c|}{Average Time (in milliseconds)}\\
\hline
	 & k=105 & k=205 & k=405 & k=605 & k=805 & k=1005  \\
\hline
$10^{2}$  & 0.54 & 0.4 & 0.69 & 0.24 & 0.31 & 0.93\\
$10^{3}$ & 2.52 & 1.63 & 2.06 & 1.61 & 1.39 & 2.68\\
$10^{4}$ & 10.29 & 8.4 & 9.96 & 14.24 & 13.65 & 15.16\\
$10^{5}$ & 71.06 & 76.49 & 114.97 & 123.29 & 160.73 & 192.22\\ 
\hline

\end{tabular}

\label{table:5}
\end{table}

For better assessing our algorithm, we run more experiments for general values of \textit{n} and \textit{k}, where we variate the size of the input array from $10^{1}$ to $10^{6}$. For each size, we generated $10^{4}$ test cases, and computed the average running time over $10^{2}, 10^{3}$ and $10^{4}$ test cases. For each input, the value of $k$ is set to the size of the array ($k=n$ for an input of size n)  which is case during combine phase  where we have arrays of size \textit{k} and we need to find \textit{k} largest elements. Results of runs of the \textbf{MAX$\_$SUM$\_$CROSS-1} procedure are shown in Table 2. As it can be seen, even for very large $n$, our algorithm takes only a few milliseconds.
Results of runs of the overall \textbf{MAX-$k$} divide and conquer algorithm are shown in Table 3 to Table 5. As it can be seen, our algorithm is very fast: for arrays of size $10^6$, with $k=10^6$, we can find the $k$ maximum subarrays in about 52 seconds.

\section{Conclusion}
In this paper, we studied the \textit{k}-maximum subarray problem and proposed a simple 
divide-and-conquer algorithm for small values of \textit{k}. Our algorithm matches the best known divide-and-conquer algorithm, while considerably simplifying the combine step.  As part of our solution, 
we provided a simple prune-and-search procedure for finding the largest $k$ values of 
$X+Y$, where $X$ and $Y$ are sorted arrays of size $n$ each. These values are computed and
stored implicitly in $O(\sqrt{k} \log^3 k)$ time, and can be reported in additional $O(k)$ time. 
Our solutions benefit from simplicity and very fast execution time, even for large values of $n$ and
$k$. We implemented our algorithms and reported excellent results.

%
%
%
\bibliographystyle{splncs04}
%

\end{document}